# Scaling Laws in Plasma Channels for Laser Wakefield Accelerators

T. L. Zhang[1], J. Y. Liu[1], S. Liu[1], R. Li[1], F. Li[1], J. F. Hua[1,a)], W. Lu[2,3,a)]

**AFFILIATIONS**

[1]Department of Engineering Physics, Tsinghua University, Beijing, China

[2]Institute of High Energy Physics, Chinese Academy of Sciences, Beijing 100049, China

[3]Beijing Academy of Quantum Information Science, Beijing 100193, China

a)E-mail: jfhua@tsinghua.edu.cn and weilu@ihep.ac.cn

**ABSTRACT**

Preformed plasma channels are essential for guiding high-power laser pulses over extended distances in laser wakefield accelerators (LWFAs), enabling the generation of multi-GeV electron beams for applications such as free-electron lasers and particle colliders. Above-threshold ionization (ATI) heating provides a robust mechanism for creating laser-matched plasma channels across a wide parameter range, owing to its density- and geometry-independent heating effect. Establishing predictive scaling laws between channel parameters and formation conditions is critical for designing channels optimized for electron acceleration across energies spanning hundreds of MeV to tens of GeV. Through combined timescale analysis and numerical simulations, hydrodynamic expansion is identified as the dominant mechanism governing density profile evolution during ATI channel formation. Remarkably, this process maintains effective laser-guiding channel structures across a wide range of initial gas density ($10^{17}$–$10^{19}$ cm$^{-3}$), as evidenced by the persistent profile similarity observed despite these significant parameter variations. For parabolic channels matched to Gaussian laser drivers, rigorous scaling laws are established that, the on-axis density scales linearly with the initial gas density, while the matching radius has an exponential dependence on both the initial gas density and the ionization laser radius. These findings provide a systematic framework for the predictive design and optimization of plasma channels in high-efficiency and high-energy LWFA applications.

## I. INTRODUCTION

Laser wakefield accelerators (LWFAs) achieve ultra-high gradients on the order of 100 GV/m, which are several orders of magnitude greater than traditional microwave-based accelerators [1], enabling dramatic reductions in the size and cost for applications such as free-electron lasers [2] and colliders [3,4]. In LWFA, for a given accelerating gradient, the energy gain scales with the acceleration distance, which is ultimately constrained by electron dephasing, laser depletion and diffraction [5,6]. While dephasing and laser energy depletion can be mitigated by optimizing the plasma density to maximize energy conversion efficiency, laser diffraction imposes a fundamental limitation. The characteristic length of diffraction (Rayleigh length $Z_R = kr^2/2$, where $k$ is the wavenumber, $r$ is the focal spot radius of the driver laser) is typically much shorter than the required accelerating distance. For example, a 200 TW laser focused to a radius of 20 μm has $Z_R \sim 1.6$ mm, far below the centimeter-scale required for GeV-level energy gain [7]. Thus, extending laser propagation well beyond the Rayleigh length across varying plasma densities and profiles is a critical challenge for high-energy LWFA.

To achieve effective laser guiding in LWFA, two principal approaches have been developed: self-focusing and preformed plasma channels. While self-focusing can occur in uniform plasmas, it typically requires petawatt-level laser powers and faces inherent instabilities due to nonlinear power dependencies [8-11]. In contrast, preformed plasma channels offer more stable and controllable guiding, making them particularly attractive for high-energy acceleration applications [12,17]. Among channel formation methods, capillary discharges [13-17] and above-threshold ionization (ATI) heating [18-21] have emerged as the most prominent techniques. The ATI method generates guiding structures through laser-induced hydrodynamic expansion, creating parabolic plasma density profiles within nanoseconds. This method exhibits key advantages over capillary discharges: (1) the guiding profile's central density and gradient can be dynamically tuned by adjusting the driver-ionization laser temporal delay [22], (2) the heating mechanism is insensitive to initial gas density and gas source geometry [23]. Advanced implementations utilize a separate modulation laser (or the leading edge of the driver laser) to create plasma channels with radii several times larger than the matching radius by ionizing surrounding neutral gas during LWFA. This technique extends the laser power attenuation length by orders of magnitude, enabling particle acceleration to 10 GeV energies over meter-scale distances [24-28].

Critical challenge remains in developing a unified framework for plasma channel generation applicable across the broad energy spectrum from hundreds of MeV to tens of GeV. Based on the LWFA scaling laws, where energy gain scales inversely with plasma density and the plasma density scales inversely with the square of driving laser radius, achieving maximal acceleration efficiency for a specific target energy requires precise optimization of the on-axis plasma density to balance dephasing length and acceleration gradient, adjustment of the matching radius to maintain optimal acceleration structure, and preservation of density profile during meter-scale laser propagation. However, current empirical approaches lack the predictive capability to navigate this multi-dimensional parameter space efficiently, underscoring the need of generalized design principles derived from a systematic investigation of plasma channel parameter dependencies on generation conditions.

Combining timescale analysis and numerical simulations, this paper identifies hydrodynamics mechanism governing channel evolution and shows that normalized density profiles remain essentially identical under different generation conditions. Scaling laws of plasma channel parameters are derived: the matching radius depends exponentially on the initial gas density and ionization laser radius, while the central density scales linearly with initial gas density. Simulation results, investigated over a wide parameter space, validated these scaling laws, enabling predictive channel design for electron acceleration from hundreds of MeV to tens of GeV.

## II. MULTIPHYSICS MODELING AND SCALING LAWS OF PLASMA CHANNEL FORMATION

### A. Simulation Framework

Channel formation process begins when an ionization laser ionizes and heats the gas, generating an outward-propagating shock wave. The gas density redistributes as the shock propagates, eventually forming a channel-shape structure. Plasma channel suitable for long-distance laser guiding is formed after being re-ionized by modulation laser, wherein the electron density becomes proportional to the gas density. Consequently, central electron density and matching radius are primarily determined by the gas density profile after shock diffusion.

To establish the relationship between plasma channel parameters and generation conditions, numerical simulations of three key stages during the channel formation process were conducted as illustrated in Fig. 1(a).

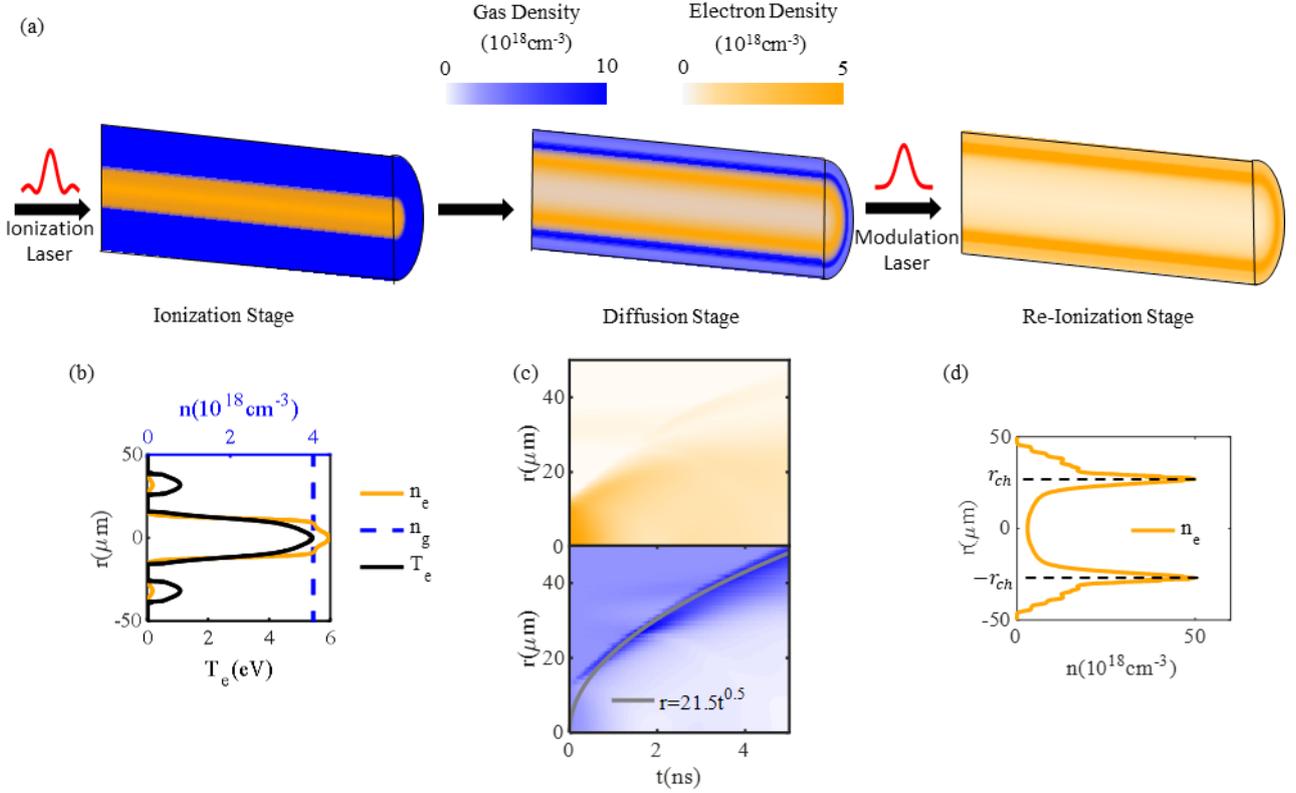

**FIG. 1.** Stages of channel formation process with particle density evolution. (a) Schematic illustration of ionization, diffusion, and re-ionization stages. Color scale: electron density (orange), gas density (blue). (b–d) Corresponding density profile (b) after Ionization, (c) during diffusion and (d) after re-ionization. The grey curve in (c) represents the fitted shock position as a function of time. Simulation parameters: Gas: nitrogen with initial density of $4\times10^{18}$ cm$^{-3}$. Ionization Laser: Bessel beam with central intensity of $8.56\times10^{14}$ W/cm$^2$ and a radius (first zero position) of 20 μm. Modulation Laser: Gaussian beam with a=1.5 and r=20 μm.

(1) Ionization stage: The ADK model was used to calculate distribution of the ionization level and electron temperature according to transverse laser intensity profile [29]. Electron thermal equilibrium was assumed due to sub-picosecond electron-electron collision timescales. Figure 1(b) presents the ionization results of the Bessel beam, whose transverse intensity distribution remains invariant during propagation. The first ring underwent weak ionization whereas the first electron of nitrogen in the central spot region was fully ionized.

(2) Diffusion Stage: Using ionization results as inputs, the evolution of density for all gas components in diffusion stage was simulated using multi-physics hydrodynamics code FLASH [30], incorporating processes including hydrodynamics, thermal conduction, heat exchange between species, ionization and recombination. Given the axisymmetric laser profile and uniform initial density, the simulations used a cylindrical coordinate. Quantities relevant to different physical mechanisms, including total electron number, electron and ion temperatures, and shock front position, were tracked to identify the dominant mechanism and to characterize the gas density evolution. Scaling laws of channel parameters were then derived based on the variation patterns of the density distribution under different initial conditions. Fig. 1(c) shows the evolution of electron (above) and gas density (below). Shock waves formed and expanded outward in the gas, with their position scaling with the square root of time, while the gas density inside gradually decreased, eventually leading to the development of a channel structure. Meanwhile, the outer ionized region also expanded but exhibited minimal impact on the density structure.

(3) Re-ionization Stage: As shown in Fig. 1(d), the entire

channel structure was fully ionized by the modulation laser, forming a plasma channel with density proportional to the original gas density. Outside this channel, the electron density followed the laser intensity profile. Simulated plasma channel parameters were then extracted from the electron density profile after re-ionization. By systematically varying the initial conditions within the target parameter range, the simulated channel parameters were fitted as functions of these initial conditions. Finally, these fitted relationships were compared with the predictions from the derived scaling laws.

**B. Theoretical Analysis of Diffusion Process**

Simulations of diffusion stage revealed similar density profiles across various initial conditions in nanosecond timescales, indicating the presence of a dominant physical mechanism. By theoretical analyzation, the following sections demonstrate that electron-ion thermalization and collisional ionization occur on significantly shorter timescales (hundreds of picoseconds) than hydrodynamics induced density evolution (nanoseconds). Additionally, normalized density profiles from all cases collapse onto a universal curve, as demonstrated in numerical simulations. These findings prove that fluid motion is the dominant process governing density evolution, thus resulting in similar outcomes. Building upon this result, an exponential relationship is derived, linking channel parameters to generation conditions for parabolic channels matching Gaussian drive beams.

To cover most applications, the parameter space is characterized by a density range of $10^{17}$ to $10^{19}$ cm$^{-3}$ and laser radius of tens of micrometers. The initial electron temperature is set to the typical temperatures of nitrogen and helium after ionization, which are 10 eV and 30 eV, respectively.

Electron-ion thermal equilibration initiates at the onset of the diffusion phase. Here, laser-heated electrons, possessing significantly higher temperatures than the initially cold ions, transfer energy to the ions via collisions. In the parameter space of interest, the characteristic time $\tau_{ei}$, is estimated to be less than 100 picoseconds according to Eq. (1):

$$\tau_{ei} = \frac{3k_B^{\frac{3}{2}}}{8\sqrt{2\pi}e^4}\frac{(m_e T_i + m_i T_e)^{\frac{3}{2}}}{(m_e m_i)^{\frac{1}{2}}Z^2 n_i \ln\Lambda_{ei}}, \quad (1)$$

where $m_e$ and $m_i$ are the electron and ion masses, $T_e$ and $T_i$ the respective temperatures, $n_e$ and $n_i$ the respective density, $Z$ the ionization state, and $\ln\Lambda_{ei}$ is coulomb logarithm [31].

Simultaneously, these hot electrons undergo collisional ionization with surrounding neutrals or ions. During ionization, energy is transferred from the primary electrons to the newly freed electrons to overcome the binding energy $E_\infty$. This energy loss causes the electron temperature to decrease and eventually leads to the cessation of collisional ionization. The characteristic timescale of ionization can be estimated through the product of the collisional ionization rate $S$ and $E_\infty$, and is on the order of hundreds of picoseconds according to Eq. (2) [32].

$$\begin{cases} \dfrac{d(n_{ele}T_{ele})}{dt} \approx -E_\infty S n_e n_i \\ S = 10^{-5}\dfrac{(T_e/E_\infty)^{1/2}}{E_\infty^{3/2}(6.0 + T_e/E_\infty)}\exp(-E_\infty/T_e) \text{ cm}^3/\text{s} \end{cases}. \quad (2)$$

Shock wave propagation is driven by the pressure gradient arising from the laser-induced electron temperature gradient. The shock formation time and its subsequent propagation speed are governed by the ion-acoustic speed $c_s = (Zk_B T_e/m_i)^{1/2}$, which calculations indicate is approximately 10 μm/ns. Given that the initial ionized region is on the order of 10 μm, the hydrodynamic evolution driven by the shock wave, which alters the density distribution to form the channel, occurs on nanosecond timescales.

**C. Numerical Analysis of Diffusion Process**

The timescale analysis was validated through numerical simulation of diffusion process. Firstly, physical quantities including electron and ion temperatures, total electron number, and shock wave position are tracked, corresponding to thermal equilibration, collisional ionization and hydrodynamic propagation, respectively. Figure 2(a) illustrate that electron-ion temperature equilibration is achieved within 90 ps, and the total electron number rises significantly only within 1 ns and subsequently remains approximately constant, showed in Fig.

2(b). The shock front propagation speed, derived from its position evolution in Fig. 1(c), closely matches the estimated 10 μm/ns. This comparison showed that, the picosecond-scale equilibration and ionization processes exert their primary influence during the initial shock formation phase, while the subsequent nanosecond-scale hydrodynamic expansion dominates the evolution of the density profile essential for channel formation.

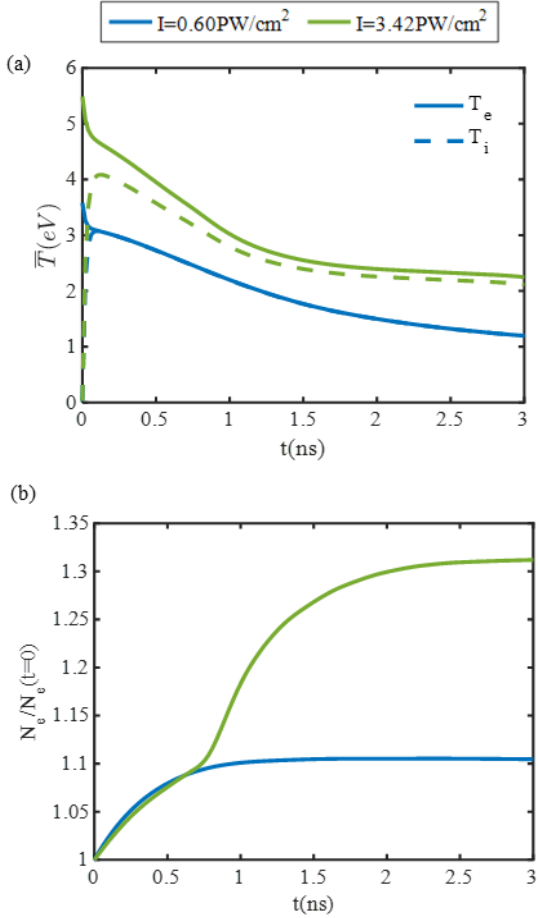

FIG. 2. Key quantities evolution under different initial ionization laser intensities. (a) Electron temperature $T_e$ (solid lines) and ion temperature $T_i$ (dashed lines). (b) Total electron number normalized to initial value

The dominance of the hydrodynamics mechanism was further investigated by examining the similarity in density evolution under different initial conditions. Gas density profiles were compared after scaling both radius and density according to the Sedov-Taylor hydrodynamic shock solution [33]: density was scaled by the initial gas density $n_{g0}$, and radius was scaled by the square root of ionization laser radius $r_0$.

Figure 3 demonstrates that the normalized density profiles evolve identically on nanosecond timescales, regardless of the specific initial values of $n_{g0}$ or $r_0$. This observed invariance confirmed that hydrodynamics governs the density evolution during the diffusion process in channel formation, solidifying its dominant role.

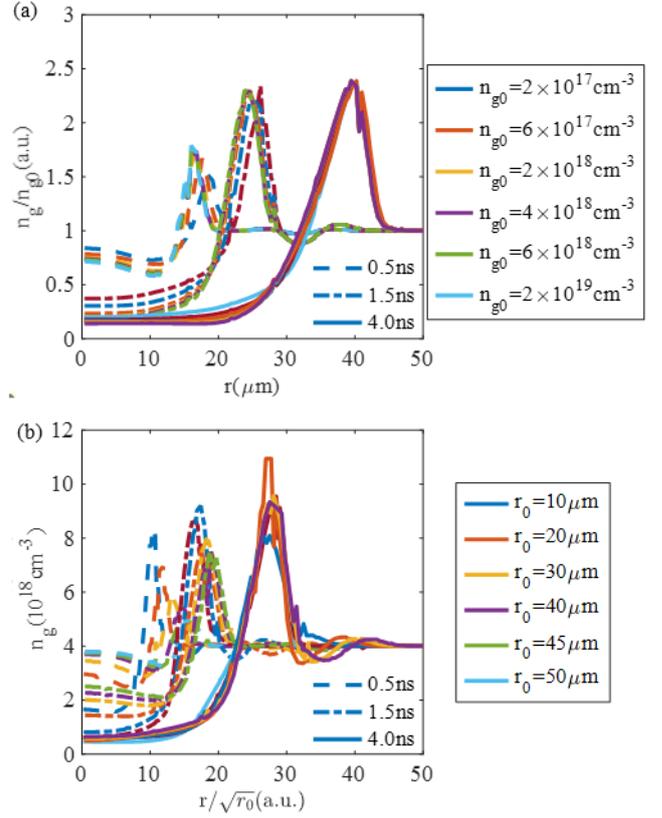

**FIG 3.** Gas density evolution during diffusion under varied initial conditions at 0.5, 1.5, 4.0 ns. (a) Density profiles at different initial densities. (b) Density profiles at different initial laser radius.

Building on the hydrodynamic dominance in density evolution established, we applied hydrodynamic normalization on density and radial coordinate by initial gas density $n_{g0}$ and ionization laser radius $r_0$. Considering coupling efficiency, parabolic plasma channels were investigated for Gaussian driver lasers [34]. In the case of parabolic density profile $n_e(r) = n_{e,a} + c_2 r^2$, scaling laws were derived for matching radius $r_w = (\pi r_e c_2)^{-0.25}$ and central density $n_{e,a}$ as:

$$\begin{cases} r_w \propto n_{g0}^{-0.25} r_0^{0.5} \\ n_{g,a} \propto n_{g0} \end{cases}, \quad (3)$$

where $r_e$ is the classical electron radius and $c_2$ is a coefficient. Additionally, the central electron density $n_{e,a}$ scales linearly with $n_{g0}$, while the channel radius $r_{ch}$ scales with $r_0$. Simulation results quantitatively validate these scaling laws through the fitting results for both nitrogen and helium gases presented in Fig. 4. Due to helium's higher ion-acoustic speed, its density evolution, which is influenced by ionization and other processes, exhibits a slight deviation in the exponent term of the relationship between matching radius and initial gas density from the value of -0.25.

These validated scaling laws establish a robust predictive framework for designing plasma channels across a broad parameter space. By applying the scaling laws, target channel parameters can be directly translated into required initial conditions. To demonstrate this design capability, generation conditions and simulated channel parameters are listed in Table 1, optimized to accelerate electrons to specific target energies spanning from hundreds of MeV to tens of GeV according to LWFA scaling laws. The corresponding electron density profiles of the channel obtained from simulations for cases 1–3 are shown in Fig. 5.

**Table 1.** Generation conditions and channel parameters for different target electron energies. The gas type is helium. $E_{ionize}$: Energy of ionization laser, $L_{ch}$: Length of channel, $E_{gain}$: Energy gain

|  | Case 1 | Case 2 | Case 3 |
| --- | --- | --- | --- |
| $r_0/\mu m$ | 10.6 | 32.8 | 40.6 |
| $n_{g0}/10^{17} cm^{-3}$ | 18.9 | 1.94 | 1.15 |
| $E_{ionize}/mJ$ | 1.5 | 50 | 120 |
| $n_{e,a}/10^{17} cm^{-3}$ | 15 | 1.5 | 0.92 |
| $r_w/\mu m$ | 11 | 37 | 47 |
| $L_{ch}/cm$ | 0.4 | 14 | 30 |
| $E_{gain}/GeV$ | 0.6 | 6 | 10 |

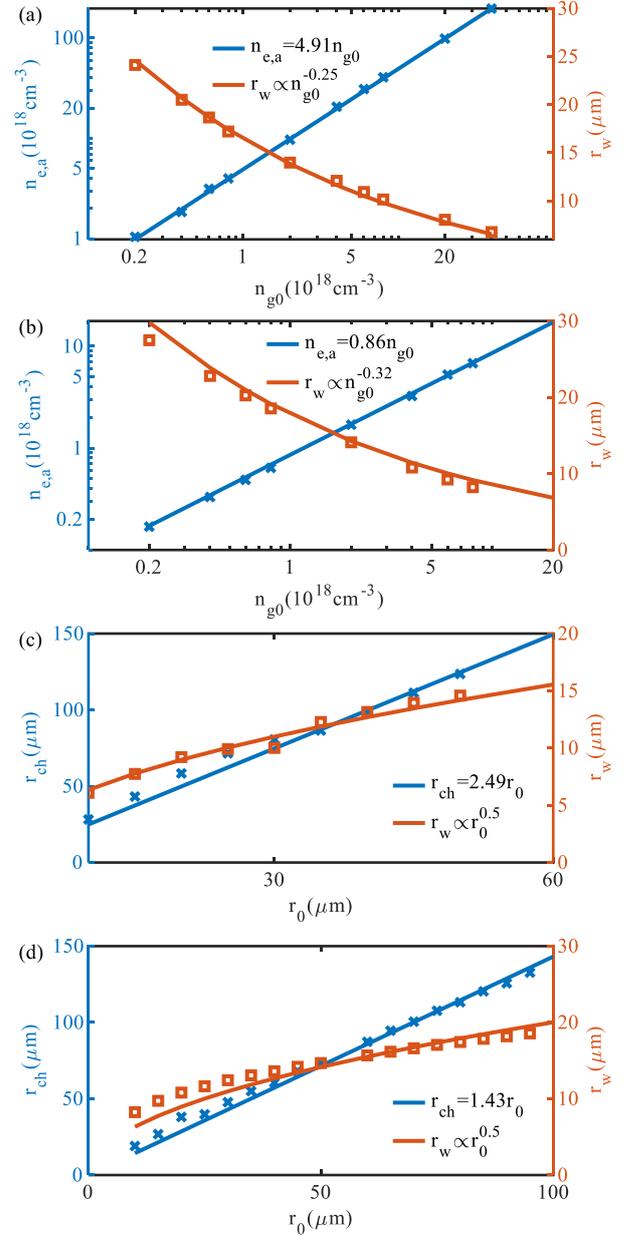

**FIG. 4.** Validation of scaling laws for plasma channel parameters: (a, b) Central density $n_{e,a}$ and matching radius $r_w$ versus initial gas density dependence for (a) nitrogen and (b) helium. (c, d) Channel half-width $r_{ch}$ and matching radius $r_w$ versus ionization laser radius $r_0$ dependence for (c) nitrogen and (d) helium. Simulation data are shown as symbols; theoretical fits as lines.

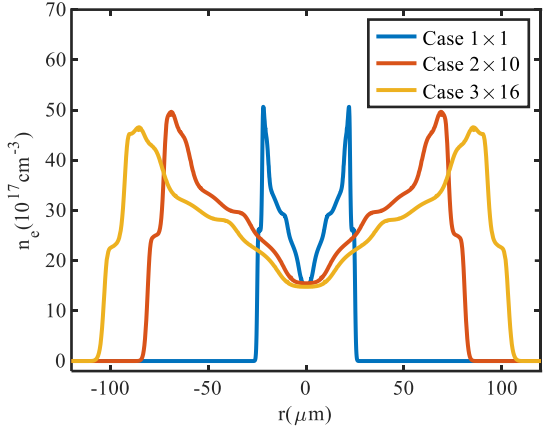

**FIG. 5.** Electron density distribution of plasma channels for Case 1, Case 2, and Case 3 in Table 1. The density values are scaled by factors of 1, 10, and 16 for Case 1, Case 2, and Case 3, respectively.

## III. CONCLUSION

In conclusion, this study established scaling laws between channel parameters and formation conditions, aiming to provide a scalable channel design methodology for LWFA. Focusing on the diffusion stage that determines channel parameters, timescales analysis shows that hydrodynamic mechanisms dominates the density profile evolution, producing similar channels under different initial conditions. Based on this result, scaling laws were derived for parabolic channels matching Gaussian drivers: the matching radius is proportional to the square root of the ionization laser radius and inversely proportional to the fourth root of the gas density, and the central density is only proportional to the gas density. The results were validated by simulations across parameter ranges spanning several orders of magnitude.

The findings hold significant value for experimental design and laser propagation optimization. Firstly, this study enables targeted on-axis density and matching radius design for diverse applications—from free-electron lasers to few-cycle laser-plasma interactions. Combined with existing research on time-domain channel evolution, this work paves the way for the design of high-quality electron acceleration. Secondly, given the negligible pressure gradient along the laser axis compared to transverse directions, diffusion processes at different axial positions operate independently. Under this approximation, our scaling laws can predict and evaluate the impact of variations in gas and ionization laser parameters on central density and matched spot size, which are critical for long-distance laser propagation.

Building on the theoretical analysis and simulation results presented in this paper, future work may experimentally characterize density evolution during plasma channel formation under varying initial conditions. Current experimental techniques lack sufficient accuracy for measuring neutral density at low densities ($\sim 10^{17}$ cm$^{-3}$), particularly in the central region, which is critical for channel parameters [28,35]. Advancements in density measurement technology are expected to enable accurate low-density channel measurements, thereby establishing new experimental benchmarks for plasma channel applications.

## ACKNOWLEDGMENTS

The authors acknowledge partial support by The Strategic Priority Research Program of the Chinese Academy of Sciences (XDB0530000), Discipline Construction Foundation of "Double World-class Project", and National Natural Science Foundation of China (Grant No. 12405169). The simulations in this paper were supported by the High Performance Computing Center of Tsinghua University.

## AUTHOR DECLARATIONS

**Conflict of Interest**

The authors have no conflicts to disclose.

**Author Contributions**

**T. L. Zhang:** Data curation, Formal analysis, Investigation, Visualization, Methodology, Software, Validation, Writing – original draft, Writing - review & editing. **J. Y. Liu:** Investigation, Writing - review & editing. **S. Liu:** Writing - review & editing. **R. Li:** Validation, Writing - review & editing. **F. Li:** Writing – review & editing. **J. F. Hua:** Conceptualization, Funding acquisition, Project administration, Resources, Supervision, Writing - review & editing, Resources. **W. Lu:** Conceptualization, Funding acquisition, Project administration, Resources, Supervision, Writing - review & editing

**DATA AVAILAILITY**

The data that support the findings of this study are available from the corresponding author upon reasonable request.

**REFEERENCES**